# Optimizing Uterine Synchronization Analysis in Pregnancy and Labor through Window Selection and Node Optimization

Kamil Bader El Dine[1], Noujoud Nader[2,3], Mohamad Khalil[3], Catherine Marque[1]

[1]CNRS UMR 7338, BMBI Sorbonne University, Université de technologie de Compiègne, Compiègne, France

[2]Louisiana State University, Baton Rouge, LA, USA

[3]Faculty of engineering, Azm center for research in biotechnology, Lebanese University

ARTICLE INFO

*Keywords:*

Preterm Labor

Windowing Approach

Connectivity Methods

Graph Theory

Neural Network

ABSTRACT

Preterm labor (PL) has globally become the leading cause of death in children under the age of 5 years. To address this problem, this paper will provide a new approach by analyzing the electrohysterographic (EHG) signals, which are recorded on the mother's abdomen during labor and pregnancy. The EHG signal reflects the electrical activity that induces the mechanical contraction of the myometrium. Because EHGs are known to be non-stationary signals, and because we anticipate connectivity to alter during contraction (due to electrical diffusion and the mechanotransduction process), we applied the windowing approach on real signals to help us identify the best windows and the best nodes with the most significant data to be used for classification. The suggested pipeline includes: i) divide the 16 EHG signals that are recorded from the abdomen of pregnant women in N windows; ii) apply the connectivity matrices on each window; iii) apply the Graph theory-based measures on the connectivity matrices on each window; iv) apply the consensus Matrix on each window in order to retrieve the best windows and the best nodes. Following that, several neural network and machine learning methods are applied to the best windows and best nodes to categorize pregnancy and labor contractions, based on the different input parameters (connectivity method alone, connectivity method plus graph parameters, best nodes, all nodes, best windows, all windows). Results showed that the best nodes are nodes 8, 9, 10, 11, and 12; while the best windows are 2, 4, and 5. The classification results obtained by using only these best nodes are better than when using the whole nodes. The results are always better when using the full burst, whatever the chosen nodes. Thus, the windowing approach proved to be an innovative technique that can improve the differentiation between labor and pregnancy EHG signals.

# 1. Introduction

Preterm labor (PL) is defined as the delivery of babies before reaching 37 weeks of gestation. Annually, over 15 million infants are impacted by PL, making it a leading contributor to child mortality in the under-five age group with over one million fatalities per year. More than 5% of preterm deliveries occur before 28 weeks of gestation, 15% occur between 28 and 31 weeks, 20% occur between 32 and 33 weeks, while the majority, 60 to 70%, occur between 34 and 36 weeks [1]. Preterm births accounted for 11.1% of all live births globally in 2010 [2]. In the United States alone, the rate of preterm birth in 2014 was between 12 and 13%; while in Europe, the recorded rate is between 5 and 9%. However, this rate exceeds 12% in low-income countries [1]. Given the global prevalence of preterm labor and its profound impact on child mortality rates, exploring effective methods for its detection becomes imperative. In this context, one of the promising avenues is to use the electrohysterogram (EHG).

The electrohysterogram (EHG), which records the electrical activity of the uterus on the mother's abdomen, is one of the promising methods for detecting PL. EHG is suggested due to its cost-effectiveness and the simplicity of equipment needed to noninvasively record uterine activity [3]. The recorded EHG reflects the electrical activity generated by active uterine muscle cells, along with the allied noise from electrical and mechanical activities. Despite this, EHG analysis has demonstrated considerable promise as one of the most effective methods for monitoring the proficiency of uterine contractions during pregnancy [4].

In the processing of EHG signals, numerous studies have employed varied concepts to distinguish between contractions during labor and pregnancy [5]. These studies have utilized diverse features capable of representing the two phenomena associated with uterine efficiency: either the cell excitability or uterine synchronization. This synchronization occurs due to heightened connectivity midst myometrial cells, characterized by the emergence of Gap Junctions and local diffusion [3]. Additionally, there is an increase in long-distance synchronization resulting from a mechanotransduction procedure [6]. For this purpose, the analysis of the Correlation/Connectivity between EHG

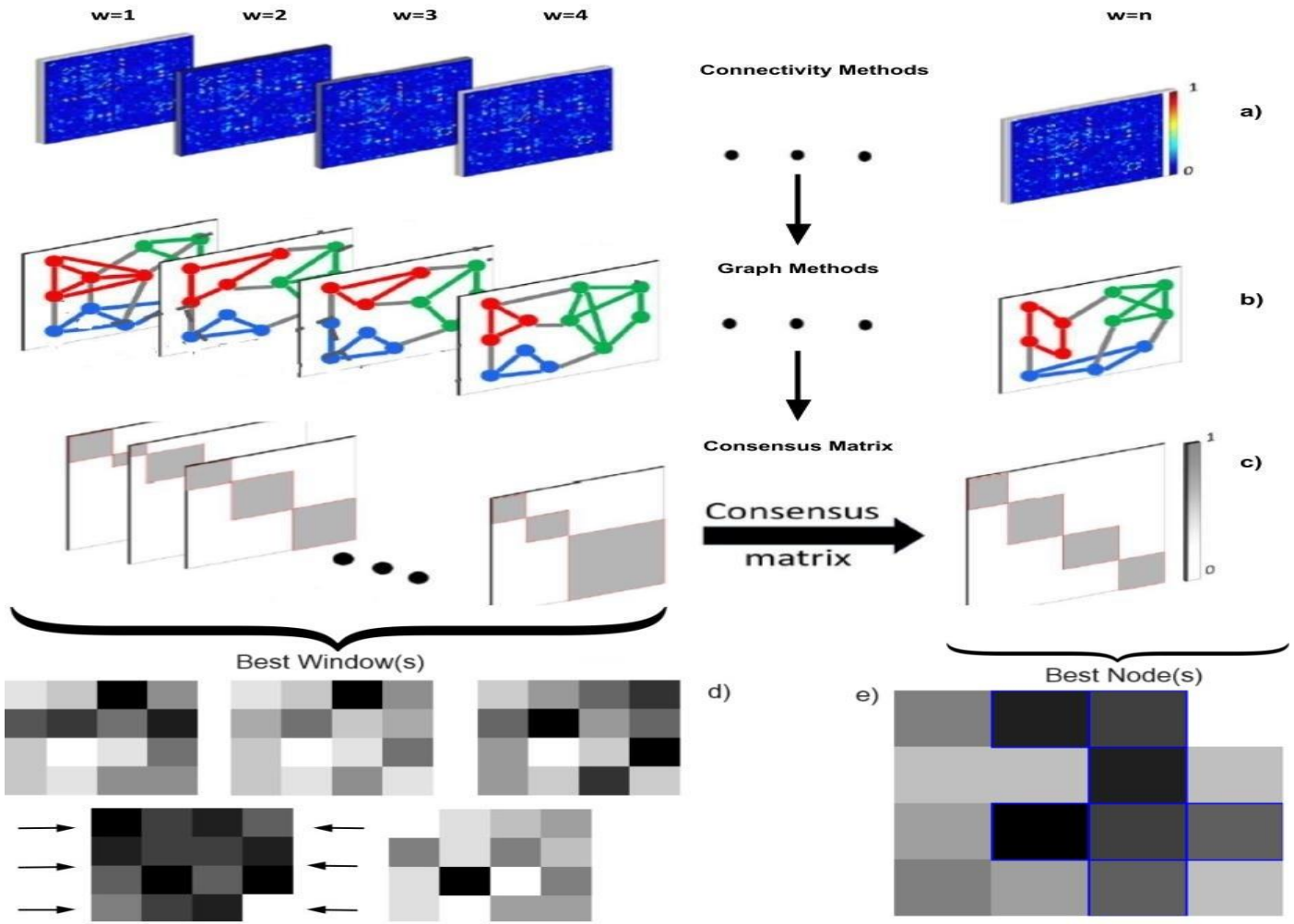

*Figure 1: The implemented approach, involving the division of signals into n windows. In (a), the estimation of the connectivity matrix is conducted for each window. (b) the extraction of Graph parameters for each window. (c), the consensus matrix is computed for each window. (d) to retrieve the best window(s) for each computation. (e) retrieve the best node(s) for each computation*

signals is one important aspect to investigate.

Numerous studies have employed diverse methodologies for this purpose, including the use of the nonlinear correlation coefficient (h2). This method was used to examine the correlations between 16 EHG signals acquired by using a matrix of 4x4 electrodes [4]. A substantial difference between labor and pregnancy EHG bursts has been evidenced [4]. EHG has also been employed to calculate the propagation velocity of electrical activity, known as conduction velocity (CV) [5]. The use of the peak frequency (PF) and propagation velocity (PV) simultaneously provided the highest performance for distinguishing between labor and non-labor EHGs [5].

Conversely, an innovative approach was introduced, employing graph theory analysis in conjunction with connectivity methods to explore the correlation among uterine electrical activities [7]. This approach aimed to leverage graph parameters to initially characterize the evolution of uterine connectivity from pregnancy to labor. Subsequently, it sought to differentiate contractions between those occurring during pregnancy and labor. Hence, the connectivity matrices were treated as graphs comprising a set of nodes (representing electrodes) connected by edges (indicating connectivity values between electrodes). The findings from this study revealed an augmentation in EHG connectivity from the pregnancy phase to the labor phase [7].

Furthermore, various approaches have been suggested for classification, with one notable method relying on Artificial Neural Networks (ANN). In this approach, measures extracted from connectivity and graph parameters are utilized as inputs to enhance the classification between labor and pregnancy [8].

The original contribution of this work is the application, for the first time on real signals, of a windowing approach in order to identify the best windows and the best nodes so as to improve the differentiation between labor and pregnancy EHG signals.

## 2. MATERIALS AND METHODS

### 2.1. Process Pipeline

In order to comprehensively analyze EHG bursts, we present a systematic process pipeline in Figure 1, which involves several key stages. We first used windowing strategy by separating each EHG burst into n windows. The correlation matrices were then estimated for each window by using different connectivity methods (Figure 1.a). The resultant connectivity matrices were subsequently analyzed as graphs, where nodes represent electrodes and edges represent connectivity values (Figure 1.b). These graphs thus represent various moments during the evolution of each EHG burst. The graph theory approach is then applied to each window to extract the graph characteristic. Finally, for each tested approach, we obtain a confusion matrix for all of the windows. Then, we computed the average of these matrices to provide a consensus matrix for every method (Figure 1.c). Following this, we retrieve the best window(s) that are the most efficient for each feature computation (Figure 1.d). We then identified the best node(s) that demonstrate optimal efficiency for each feature computation (Figure 1.e).

### 2.2. Connectivity Methods

We utilized four connectivity measures from the 16 EHG signals, incorporating the classical linear ($R^2$) and nonlinear ($H^2$) correlation coefficients, and the imaginary part of the coherence (ICOH), all of which have been shown to be promising in previous studies [3].

**The cross-correlation coefficient ($R^2$):** measures the strength of the linear correlation relationship among two variables, X and Y, within the time domain [9].

$$R^2 = max_t \frac{cov^2(X(t), Y(t+\tau))}{var(X(t))var(Y(t+\tau))} \quad (1)$$

where *cov* and var are the covariance and variance among the two-time series *X(t)* and *Y(t),* respectively. $\tau$ reflects the change in time.

**The nonlinear correlation ($H^2$)**: quantifies the nonlinear relationship between two variables. It is calculated by assessing the value of X as a function of the value of Y from the two signals, X(t) and Y(t), each of length N. A nonlinear regression curve can be employed to predict the value of Y based on X [4]. The unexplained variance is determined by subtracting the explained variance from the original variance. According to the regression curve, $H^2$, the nonlinear correlation value, reflects the decrease in Y variance achievable by forecasting Y values from X, expressed as $H^2$ = (total variance - unexplained variance)/total variance.

$$H^2_{X/Y} = \frac{\sum_{k-1}^{N} Y(k)^2 - \sum_{k-1}^{N}(Y(K) - f(X_i))^2}{\sum_{k-1}^{N} Y(k)^2} \quad (2)$$

where $f(X_i)$ is the nonlinear regression curve (linear piecewise approximation).

**Imaginary part of coherence (Icoh)**: Coherence $C_{XY}$ is a metric that has been widely used to demonstrate, in the frequency domain, the links among two signals, X and Y, as a function of the frequency [10], where volume conduction has a direct influence on the genuine coherence value. When electrical activity is collected and processed at a distance from its source, such as when monitoring abdominal EHGs, an effect of the volume conduction occurs. As a result, new solutions for addressing this issue have been proposed, focused solely on the imagined component of the coherence. The basic notion is that an interaction with zero-lag of the real sections of the coherence function between signals suggests a false interaction. But the imaginary component of the coherence function may reveal authentic interactions, indicating true signal correlation.

$$ICOH = \frac{|\text{Im}C_{XY}(f)|}{\sqrt{|C_{XX}(f)||C_{YY}(f)|}} \quad (3)$$

where the coherence (C) function offers the linear correlation between two signals, X and Y, as a function of frequency. It is determined by normalizing their respective auto-spectral density functions, $C_{XX}$ and $C_{YY}$, with their cross-spectral density function, $C_{XY}$.

### 2.3. Graph Theory

The estimated connectivity matrices obtained at the previous step are then analyzed as graphs. A graph is a mathematical abstract structure composed of vertices (V) or nodes, and edges (E) that link pairs of those vertices. In our study, nodes correspond to the electrodes and edges correspond to the calculated connectivity values [11]. Furthermore, for each corresponding graph, five graph parameters were extracted: Strength (Str), Clustering Coefficient (CC), Efficiency (Eff), PageRank (PR), and Betweenness Centrality (BC).

**Strength (Str):** of a node indicates its importance and connectivity in respect to other nodes in the network. A node strength is the sum of the weights of the edges connecting to it.

$$S_i = \sum_{j \in N} w_{ij} \quad (4)$$

where *i* and *j* represent the $i_{th}$ and $j_{th}$ nodes, correspondingly. *N* is the total number of nodes in the graph, and $w_{ij}$ denotes the connectivity value for the relationship between *i* and *j* [12].

**Clustering Coefficient (CC):** indicates the extent to which nodes frequently interact or connect to one another. It represents the degree to how a particular node's neighbors link to each other.

$$C_i = \frac{2t_i}{k_i(k_i - 1)} \quad (5)$$

Where *i* is the node, $t_i$ is the number of triangular connections between nodes, and $k_i(k_i-1)$ is the number of maximum possible edges in the network [13].

**Efficiency (Eff):** shows a proxy measure of graph clustering characteristics [14]. It is the inverse of the shortest path between two nodes

$$E = \frac{1}{N(N-1)} \sum_{i,j \in N, i \neq j} \frac{1}{d_{ij}} \quad (6)$$

where *i* and *j* denote the $i_{th}$ and $j_{th}$ nodes respectively. The shortest path between two nodes *i* and *j* is represented by the value $d_{ij}$. N represents the total number of nodes in the network.

**PageRank (PR):** the PageRank algorithm still operates on a network of nodes and edges. The importance or influence of a node is calculated based on the contributions from other nodes connected to it, taking into account a damping factor *d*, to model the probability of following links or randomly transitioning to other nodes. The PR value is the number of links pointing to a given node [15].

$$PR(u) = (1 - d) + d \sum_{U \in Bu} \frac{PR(u)}{N_u} \quad (7)$$

where *u* represents the node (electrode), *Nu* the number of connections from *u*, and *d* the damping factor, which can range between 0 and 1.

**Betweenness Centrality (BC):** Betweenness centrality is concerned with locating nodes that are frequently encountered on the shortest path between two other nodes [16]. As a consequence, betweenness centrality creates a relational value based on a node local role with respect to the nodes in between [17]. Nodes identified on a path between two other nodes manage the flow of information between them, with a complete control (when only a singular path can be present between the two further nodes) or a restricted control (when several pathways exist between nodes [16]. It keeps track of how many times a node is placed on the shortest path between other nodes. It determines how well the studied node can function as a communication control point.

$$BC(v) = \sum_{s \neq v \neq t} \frac{\sigma st(v)}{\sigma_{st}} \quad (8)$$

where, σst(v) denotes the count of shortest paths passing through vertex v along the trail from s to t, while σst represents the count of shortest paths from s to t [18].

### 2.4. Consensus Matrix

The consensus matrix serves as a tool for identifying communities within extensive networks [19], such as EHG bursts. It assigns values indicating the relative strength of connections between each node concerning how much more tightly connected they are with other nodes in a real network compared to their connections in a random network. Subsequently, the process merges communities into a single node and applies modularity clustering on the condensed graphs using the Louvain algorithm [20]. This yields the most crucial nodes within the network. Finally, a final consensus matrix is generated by calculating the ratio of each node to the other nodes in the same module across all time frames, employing the same approach as before.

*2.5. Machine learning Approach*

Throughout this study, two machine learning approaches have been proposed in order to determine the strategy that best classified labor and pregnancy contractions: Logistic Regression (LR) and Multilayer perceptron (MLP). We chose to test these two approaches as previous studies showed that LR and MLP were the best methods for the classification between labor and pregnancy EHG bursts [21].

**Logistic Regression (LR):** it can be employed when the research approach is focused on identifying whether or not an event occurred (no time course information is needed). It is extensively used in health sciences research [14] and is especially beneficial for binary models as sickness states (diseased or well) and decision-making (yes or no). The logistic regression model is founded on the function of logistic that predicts and defines the connection between a dependent variable Y. The output Y has just two potential values, which result from the existence or absence of an event, as well as independent factors that influence that phenomena.

$$P = \frac{e^{(b0+b1*x)}}{1+e^{(b0+b1*x)}} \quad (9)$$

where weights or coefficient values are represented by *b0* and *b1*. The bias or intercept is represented by *b0*, while the coefficient is represented by *b1*.

**The Multilayer Perceptrons (MLP)** model appears to be the most often used artificial neural network (ANN) type for data modeling. MLP network design consists of neurons stacked in layers (Input Layer, Hidden Layer(s), Output Layer) [22]. The MLP model is a feedforward neural network, which is a form of ANN. MLP is a supervised classification approach. A feedforward neural network is a fundamental type of neural network that can imitate continuous functions.

The following two equations can be used to describe a neuron K in MLP network:

$$y_k = f(u_k + b_k) \quad (11)$$

$$u_k = \sum_{i=1}^{n} w_{ki} x_i \quad (12)$$

where $x_1, x_2, ..., x_n$ are the input signals, $w_{k1}, w_{k2}, ..., w_{kn}$ are the neuron connection weights, $u_k$ represents the linear output resulting from the weighted combination of inputs, where $b_k$ is the bias term, *f* is the activation function, and $y_k$ signifies the output signal of the neuron.

As our results first showed that the best results are obtained with logistic regression, we used it in the final phases. Its key benefit is that it eliminates distracting effects by studying the fluctuation of all variables simultaneously [23].

*2.6. Data*

To capture the EHG signals, we used the 16 monopolar electrodes of a 4x4 matrix positioned on the mother's belly. The grid was positioned so that the third column of electrodes corresponded with the uterine median vertical axis and the 10-11h electrodes were put in the center of the uterus [24]. The signals are digitized at a sample rate of 200 hertz. The EHG signals are afterward manually segmented and denoised with the CCA-EMD technique [25]. As a result, we obtained 247 pregnant and 183 labor contractions after splitting and denoising. The following analysis is performed on these resultant contractions.

## 3. RESULTS

As mentioned previously, the first step in this study is to segment the whole contractions into windows. For this aim, it is necessary to establish the window length and the number of windows to be used to represent each EHG contraction. However, to compare the results of all the contractions, one must use the same length and number of windows to investigate them. As a result, all signals in our study should have the same set duration.

We examined alternative signals lengths to select a minimum common length of analysis for all the signals and we came up with a period of 60 seconds (12000 points) to represent each contraction. We used the power approach [26] to determine the position of this 60-second period along each contraction. To do so, we determined the maximum of each EHG power to pinpoint the signal portion with the greatest power. We then chose the 60 seconds of interest for the windowing strategy by taking 30 seconds before and 30 seconds after this maximum power. The length of the resulting signal is thus 12000 points (60 seconds). As a result of this pre-processing, we standardized all signals to the same length, allowing for an equal division into windows of uniform length.

As for the window length, we chose a duration of 3000 points (15 seconds) and 50% overlapping windows (Figure 2). Indeed, these values were previously used by [27] to quantify the instantaneous phase difference of instantaneous amplitude correlation. As each signal has 12000 points, we get 7 windows for each EHG signal.

As previously stated, all processed EHG signals have the same duration and are separated into seven windows. For each window, the connectivity methods, graph methods, and consensus matrix were calculated. Finally, for every method window, a consensus matrix will be estimates.

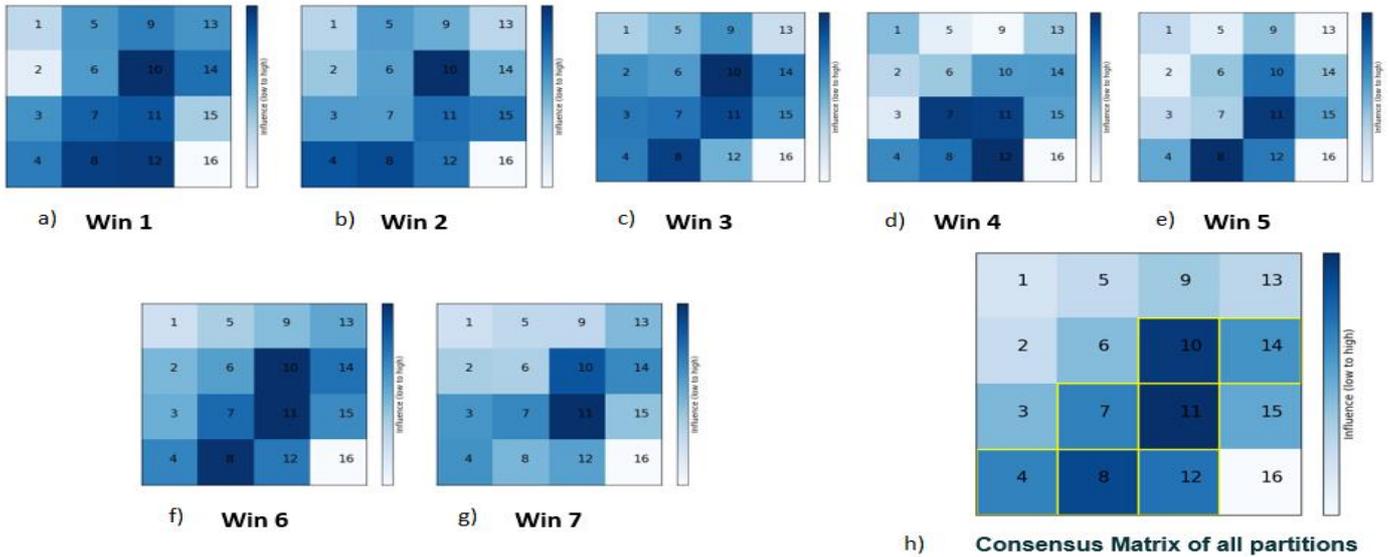

Figure 3: Consensus matrices of each window utilizing $R^2(BC)$: from a) to g), presenting the results of each window sequentially from window 1 to window 7, h) showing the mean consensus matrix over all windows

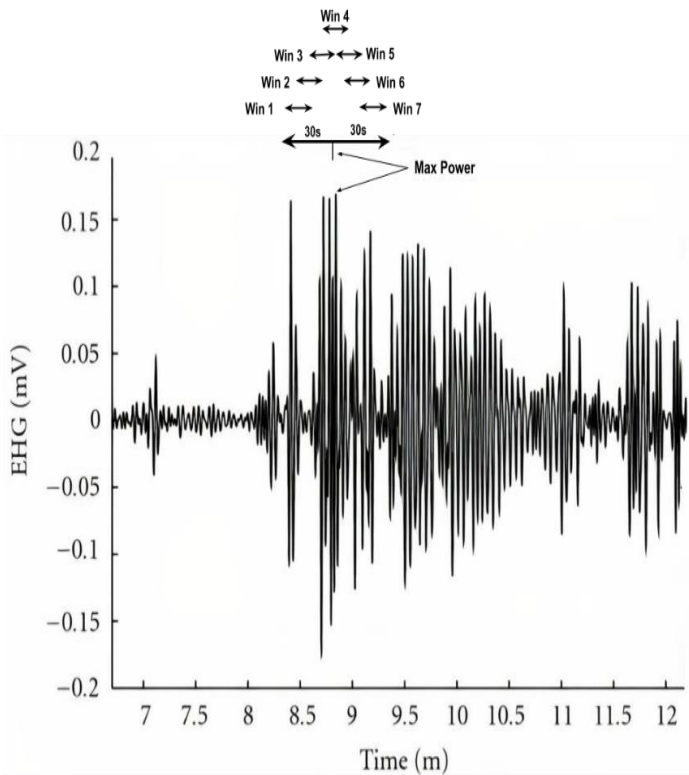

Figure 2: Example of the EHG signal, indicating the location of the maximum power energy and the position of each window

In Figure 3 (3.a-3.g) we are showing an example of the results of the consensus matrices for the 7 windows and the mean consensus matrix for all windows in Figure 3.h using $R^2(BC)$. The labels from 1 to 16 represent the nodes (electrodes) label. The consensus presents the $R^2(BC)$ value for each node with a color-based presentation: the darker the color is, the higher is the value of $R^2(BC)$. A yellow box highlights the most significant nodes in the final average consensus matrix. The selected nodes here are: 4, 7, 8, 10, 11, 12, and 14.

Following this, we calculated the $R^2(BC)$ values for the most important nodes. Figure 4(a) depicts the most critical node values for each window. Most of these nodes have the highest $R^2(BC)$ values in window 2, except for node 7, which has the greatest value in window 3. The boxplot of the most important nodes for each window is shown in Figure 4(b). Window 2 proved to get the best outcomes.

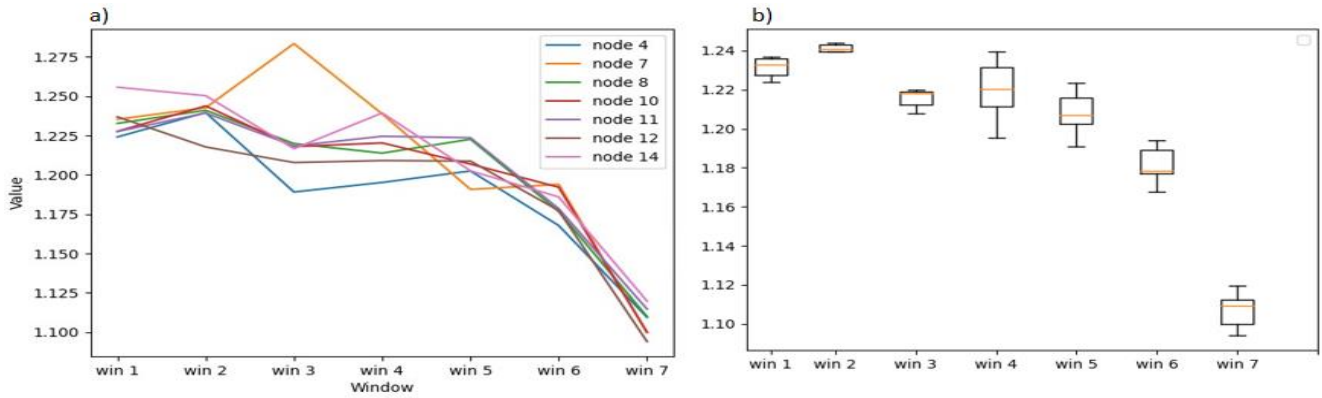

*Figure 4. Analyses for each window using the R2(BC) method. Panel a) presents the values of the most important nodes (4,7,8,10,11,12,14) in each window, panel b) displays a boxplot for the most important nodes in each window*

Table 1: **Best window(s) for each method**

| Method | Best Window(s) | Method | Best Window(s) |
|---|---|---|---|
| $R^2$ | Window 6 | $H^2(PR)$ | Window 4 |
| $R^2(Str)$ | Window 4 and Window 5 | $H^2(BC)$ | Window 7 |
| $R^2(Eff)$ | No noticeable best window | $H^2(CC)$ | Window 2 and Window 5 |
| $R^2(PR)$ | Window 4 and Window 5 | ICOH | Window 7 |
| $R^2(BC)$ | Window 3 and Window 4 | ICOH(Eff) | Window 4 |
| $R^2(CC)$ | Window 2 and window 7 | ICOH(Str) | Window 5 and Window 7 |
| $H^2$ | Window 5 | ICOH(PR) | Window 2 and Window 6 |
| $H^2(Str)$ | Window 7 | ICOH(BC) | Window 2 and Window 5 |
| $H^2(Eff)$ | Window 4 | ICOH(CC) | Window 2 and Window 4 |

Table 1 presents the results for the best window(s) across all methods and parameters.

At last, we computed the average consensus matrix of all approaches in order to demonstrate the optimal nodes, windows, and parameters from all methods for the classification of labor and pregnancy contractions. Figure 5 depicts the optimal nodes as 8, 9, 10, 11, and 12. Nodes 9-12 are located on to the mother's abdomen median vertical axis and have previously been shown to be the optimum electrode placement for recording EHG [16].

The greatest results were found in window 4 (selected 7 times), which was positioned directly in the middle of the analyzed window (justifying the choice of the maximum of power to pick the processed signal), trailed by window 5 (selected 6 times), and 2 (selected 5 times) as presented in table 1.

In addition, we will use the parameters selected in a previous study [3]. These five parameters are: ICOH(Str), ICOH(Eff), ICOH(CC), $H^2(BC)$, and $R^2(Eff)$.

For the classification step, we applied artificial intelligence tools to the best results obtained in the previous phases.

Thus, we used the previously selected best windows (2, 4, and 5), and best nodes (8, 9, 10, 11, and 12). We then computed, as input of a classifier, the best parameters (ICOH(Str), ICOH(Eff), ICOH(CC), $H^2(BC)$, $R^2(Eff)$ extracted from these selected windows and nodes, as well as from the whole signal and the whole nodes. As previously explained, we chose Logistic regression and MLP as classification methods [18]. Then, for each EHG contraction, we compared the results obtained by windowing to those obtained by combining all 7 windows and then the whole signal. We also compared the results obtained when just the best nodes were used, compared to the results obtained when all nodes were used. The measurement metric used was the Area under curve (AUC). Table 2 summarizes the results of this process.

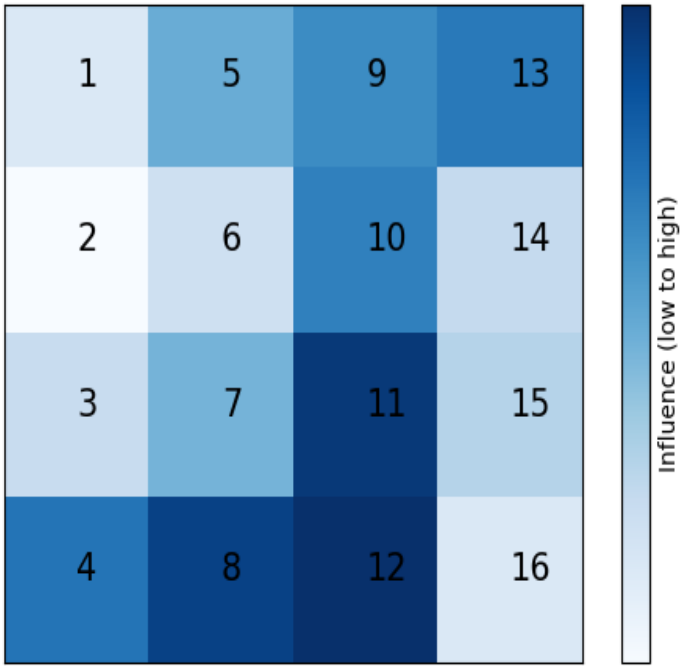

Figure 5: Average consensus matrix derived from all windows and methods

Table 2: **Results of different methods, windows and nodes**

| Method | Window | Nodes | AUC Value |
|---|---|---|---|
| Logistic Regression | Window 2 | Best nodes | 0.825 |
| MLP | Window 2 | Best nodes | 0.797 |
| Logistic Regression | Window 2 | All nodes | 0.836 |
| MLP | Window 2 | All nodes | 0.841 |
| Logistic Regression | Window 4 | Best nodes | 0.865 |
| MLP | Window 4 | Best nodes | 0.797 |
| Logistic Regression | Window 4 | All nodes | 0.838 |
| MLP | Window 4 | All nodes | 0.841 |
| Logistic Regression | Window 5 | Best nodes | 0.821 |
| MLP | Window 5 | Best nodes | 0.771 |
| Logistic Regression | Window 5 | All nodes | 0.792 |
| MLP | Window 5 | All nodes | 0.821 |
| Logistic Regression | All Windows | Best nodes | 0.911 |
| MLP | All Windows | Best nodes | 0.896 |
| Logistic Regression | All Windows | All nodes | 0.902 |
| MLP | All Windows | All nodes | 0.883 |
| **Logistic Regression** | **Whole signal** | **Best nodes** | **0.918** |
| MLP | Whole signal | Best nodes | 0.903 |
| Logistic Regression | Whole signal | All nodes | 0.914 |
| MLP | Whole signal | All nodes | 0.897 |

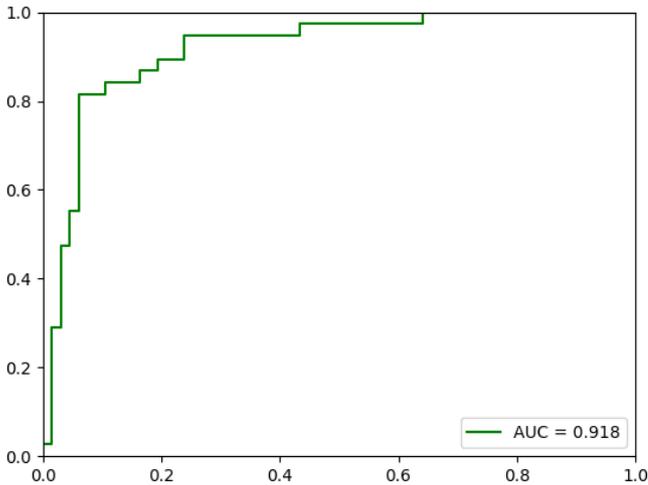

Figure 6: Whole signals - best nodes - Logistic Regression (AUC: 0.918)

As indicated in table 2, the best results were achieved when the entire signal and the best nodes were used (AUC=0.918). Then, the next best result was reached when we merge all the windows (AUC=0.911), which produced a better outcome than by using only one window. However, when using only one window, window 4 (AUC=0.865) performed better than windows 2 and 5. In figure 6 we present the roc curve of the best result, obtained when combining the features from the entire signal with the best nodes.

As a consequence, we may conclude that selecting the best nodes rather than all nodes produces better classification results. The windowing approach, nevertheless, did not enhance the classification between labor and pregnancy contractions.

In terms of window size, the AUC appears to increase as the signal duration increases. This is supported by the findings shown in Table 3. We calculated the features from the best nodes either from their unique best window(s) (row 1) or from the concatenation of the windows of interest for the majority of the features (W2, W4, and W5) (row 2). When using Logistic regression, the minimal area under curve achieved (AUC=0.865) is equivalent to or greater than when using only one window (AUC=0.821-0.865). The highest value obtained (AUC=0.881) is lower compared to when all windows (W1 to W7) are utilized together (AUC=0.902) or when the entire signal is considered (AUC=0.918).

Table 3: **Results of different windows size, with best nodes and Logistic regression**

| Condition | Window choice | AUC Value |
|---|---|---|
| Specific window(s) | ICOH(Str): window 5 and window 7 concatenated<br><br>ICOH (Eff): window 4<br><br>ICOH(CC): window 2 and window 4 concatenated<br><br>$H^2$(BC): window 7 | 0.865 |
| Best windows concatenated | Window 2, window 4 and window 5 concatenated | 0.881 |

## 4. Discussion

We established an innovative approach for identifying connectivity from EHG signals, considering the signal non-stationarity as well as the connectivity temporal evolution. In our application to real EHG signals, we implemented a windowing approach to identify and select the best window containing the most informative data.

To analyze all signals of the same duration, the EHG signals were first scaled to 12000 points (60 seconds), concentrated on the more powerful part of the EHG burst. The selected sections of the signals were then parted into 7 windows to investigate the windowing effect on classification and to determine which window delivers the best results.

We used two approaches, Logistic regression and MLP, to classify pregnancy and labor based on variables derived from connectivity methods (ICOH, $R^2$, and $H^2$) with or without graph parameters (Str, CC, Eff, PR, and BC).

Noticeably, regardless of the windows used, picking only the best nodes produces better results than using all nodes.

On the contrary, the AUC recorded for logistic regression on the entire signals and the best nodes (AUC=0.918) was marginally higher than the combination of all windows and higher than the selection of only one window (the best result with all windows and best nodes being AUC=0.911, whereas the best result with only one window and best nodes, specifically window 4, is AUC=0.865).

## 5. CONCLUSION

In this study, we established that based on the obtained data, we can deduce that the AUC values improve when we choose the best nodes rather than all 16 nodes, and they also increase as the signal size increases. As a consequence, the results are better when we use the entire contraction (the entire signal duration for each EHG contraction rather than the chosen 60s). Nonetheless, when we compare the execution timings of the two instances, we see that the execution time dropped as the number of nodes and signal length decreased. As a result, the primary challenge is whether we can lose information (approximately 0.3%) to gain on the execution time.

Finally, the windowing strategy may successfully be used to reduce the time required to classify between pregnancy and labor contractions, while simultaneously sacrificing some precision in the classification. This method might also be useful in diagnosing preterm labor early on by allowing for rapid characterization of uterine contractions and risk identification throughout pregnancy.

**Acknowledgements**

This study was supported by Partenariats Hubert Curien: PHC Cèdre 2020, project N°40178YG, University of Technology of Compiègne (UTC-France) and Lebanese University (LU-Lebanon).

**References**

[1] R. L. Goldenberg, J. F. Culhane, J. D. Iams, and R. Romero, "Epidemiology and causes of preterm birth," *Lancet*, vol. 371, no. 9606, pp. 75–84, Jan. 2008, doi: 10.1016/S0140-6736(08)60074-4.

[2] H. Blencowe *et al.*, "National, regional, and worldwide estimates of preterm birth rates in the year 2010 with time trends since 1990 for selected countries: a systematic analysis and implications," *Lancet*, vol. 379, no. 9832, pp. 2162–2172, Jun. 2012, doi: 10.1016/S0140-6736(12)60820-4.

[3] K. B. E. Dine, "Uterine synchronization analysis during pregnancy and labor using graph theory, classification based on machine learning".

[4] B. Karlsson, J. Terrien, V. Gudmundsson, T. Steingrimsdottir, and C. Marque, "Abdominal EHG on a 4 by 4 grid: mapping and presenting the propagation of uterine contractions," *11th Mediterranean Conference on Medical and Biomedical Engineering and Computing 2007*, p. 139, 2007.

[5] M. LUCOVNIK *et al.*, "Use of uterine electromyography to diagnose term and preterm labor," *Acta Obstet Gynecol Scand*, vol. 90, no. 2, pp. 150–157, Feb. 2011, doi: 10.1111/j.1600-0412.2010.01031.x.

[6] S. Mohammadi Far, M. Beiramvand, M. Shahbakhti, and P. Augustyniak, "Prediction of Preterm Labor from the Electrohysterogram

Signals Based on Different Gestational Weeks," *Sensors (Basel)*, vol. 23, no. 13, p. 5965, Jun. 2023, doi: 10.3390/s23135965.

[7] C. Rabotti, M. Mischi, J. van Laar, G. Oei, and J. Bergmans, "Inter-electrode delay estimators for electrohysterographic propagation analysis," *Physiological measurement*, vol. 30, pp. 745–61, Jul. 2009, doi: 10.1088/0967-3334/30/8/002.

[8] M. Lucovnik *et al.*, "Noninvasive uterine electromyography for prediction of preterm delivery," *Am J Obstet Gynecol*, vol. 204, no. 3, p. 228.e1–10, Mar. 2011, doi: 10.1016/j.ajog.2010.09.024.

[9] K. Ansari-Asl, F. Wendling, J. J. Bellanger, and L. Senhadji, "Comparison of two estimators of time-frequency interdependencies between nonstationary signals: application to epileptic EEG," *Conf Proc IEEE Eng Med Biol Soc*, vol. 2006, pp. 263–266, 2004, doi: 10.1109/IEMBS.2004.1403142.

[10] Z. Ali *et al.*, "Forecasting Drought Using Multilayer Perceptron Artificial Neural Network Model," *Advances in Meteorology*, vol. 2017, p. e5681308, May 2017, doi: 10.1155/2017/5681308.

[11] C. J. Nelson and S. Bonner, "Neuronal Graphs: A Graph Theory Primer for Microscopic, Functional Networks of Neurons Recorded by Calcium Imaging," *Front Neural Circuits*, vol. 15, p. 662882, Jun. 2021, doi: 10.3389/fncir.2021.662882.

[12] M. Rubinov and O. Sporns, "Complex network measures of brain connectivity: Uses and interpretations," *NeuroImage*, vol. 52, no. 3, pp. 1059–1069, Sep. 2010, doi: 10.1016/j.neuroimage.2009.10.003.

[13] M. Hassan, J. Terrien, C. Muszynski, A. Alexandersson, C. Marque, and B. Karlsson, "Better pregnancy monitoring using nonlinear correlation analysis of external uterine electromyography," *IEEE Trans Biomed Eng*, vol. 60, no. 4, pp. 1160–1166, Apr. 2013, doi: 10.1109/TBME.2012.2229279.

[14] S. Boccaletti, V. Latora, Y. Moreno, M. Chavez, and D.-U. Hwang, "Complex networks: Structure and dynamics," *Physics Reports*, vol. 424, no. 4, pp. 175–308, Feb. 2006, doi: 10.1016/j.physrep.2005.10.009.

[15] S. Ding, "Feature Selection Based F-Score and ACO Algorithm in Support Vector Machine," *Knowledge Acquisition and Modeling, International Symposium on*, vol. 1, pp. 19–23, Jan. 2009, doi: 10.1109/KAM.2009.137.

[16] L. Leydesdorff, "'Betweenness Centrality' as an Indicator of the 'Interdisciplinarity' of Scientific Journals," *Journal of the American Society for Information Science and Technology*, vol. 58, Jul. 2007, doi: 10.1002/asi.20614.

[17] L. Freeman, "A Set of Measures of Centrality Based on Betweenness," *Sociometry*, vol. 40, pp. 35–41, Mar. 1977, doi: 10.2307/3033543.

[18] A. Plutov and M. Segal, "The delta-betweenness centrality," Sep. 2013, pp. 3376–3380. doi: 10.1109/PIMRC.2013.6666731.

[19] D. S. Bassett, M. A. Porter, N. F. Wymbs, S. T. Grafton, J. M. Carlson, and P. J. Mucha, "Robust detection of dynamic community structure in networks," *Chaos*, vol. 23, no. 1, p. 013142, Mar. 2013, doi: 10.1063/1.4790830.

[20] V. Blondel, J.-L. Guillaume, R. Lambiotte, and E. Lefebvre, "Fast Unfolding of Communities in Large Networks," *Journal of Statistical Mechanics Theory and Experiment*, vol. 2008, Apr. 2008, doi: 10.1088/1742-5468/2008/10/P10008.

[21] K. B. El Dine, N. Nader, M. Khalil, and C. Marque, "Uterine Synchronization Analysis During Pregnancy and Labor Using Graph Theory, Classification Based on Neural Network and Deep Learning," *IRBM*, vol. 43, no. 5, pp. 333–339, Oct. 2022, doi: 10.1016/j.irbm.2021.09.002.

[22] A. H. Marblestone, G. Wayne, and K. P. Kording, "Toward an Integration of Deep Learning and Neuroscience," *Frontiers in Computational Neuroscience*, vol. 10, 2016, Accessed: Oct. 10, 2023. [Online]. Available: https://www.frontiersin.org/articles/10.3389/fncom.2016.00094

[23] P. Ranganathan, C. S. Pramesh, and R. Aggarwal, "Common pitfalls in statistical analysis: Logistic regression," *Perspect Clin Res*, vol. 8, no. 3, pp. 148–151, 2017, doi: 10.4103/picr.PICR_87_17.

[24] E. Y. Boateng and D. A. Abaye, "A Review of the Logistic Regression Model with Emphasis on Medical Research," *JDAIP*, vol. 07, no. 04, pp. 190–207, 2019, doi: 10.4236/jdaip.2019.74012.

[25] S. Sperandei, "Understanding logistic regression analysis," *Biochemia Medica*, vol. 24, no. 1, pp. 12–18, Feb. 2014, doi: 10.11613/BM.2014.003.

[26] L. Chen, L. Heikkinen, C. Wang, Y. Yang, E. Knott, and G. Wong, "miRToolsGallery : a tag-based and rankable microRNA bioinformatics resources database portal," *Database : The journal of Biological Databases and Curation*, vol. 2018, no. 0, 2018, doi: 10.1093/database/bay004.

[27] L. C. Freeman, "Centrality in social networks conceptual clarification," *Social Networks*, vol. 1, no. 3, pp. 215–239, Jan. 1978, doi: 10.1016/0378-8733(78)90021-7.